\documentclass[a4paper]{jpconf}

\usepackage{amsmath,amssymb,amsfonts}
\usepackage{graphicx}
\usepackage{hyperref}
\usepackage{color}
\usepackage{array}
\usepackage[normalem]{ulem}

\usepackage{fancyhdr}
\begin{document}

\title{Exploring new frontiers in statistical physics\break{} with a new, parallel Wang--Landau framework}

\author{Thomas Vogel$^{1,2}$, Ying Wai Li$^{1,3}$, Thomas W\"ust$^4$, and David~P.~Landau$^1$}

\address{$1$ Center for Simulational Physics, The University of
  Georgia, Athens, GA 30602, USA}

\address{$2$ Current address: Theoretical Division, Los Alamos
  National Laboratory, Los Alamos, NM~87545, USA}

\address{$3$ Current address: National Center for Computational
  Sciences, Oak Ridge National Laboratory, Oak Ridge, TN 37831, USA}

\address{$4$ Swiss Federal Research Institute
  WSL, Z\"{u}rcherstrasse 111, CH-8903 Birmensdorf, Switzerland}

\ead{thomasvogel@physast.uga.edu}

\begin{abstract}
\noindent
Combining traditional Wang--Landau sampling for multiple replica
systems with an exchange of densities of states between replicas, we
describe a general framework for simulations on massively parallel
Petaflop supercomputers.  The advantages and general applicability of
the method for simulations of complex systems are demonstrated for the
classical 2D Potts spin model featuring a strong first-order
transition and the self-assembly of lipid bilayers in amphiphilic
solutions in a continuous model.
\end{abstract}



\pagestyle{fancy}

\section{Introduction}

In recent years the goal of Petaflop computing has been achieved by
replying on massively parallel systems.  This development requires a
new approach to the efficient utilization of computing resources;
moreover, improvements in methodology in computational statistical
physics have also taken place.  One such improvement, known as
``Wang--Landau sampling'', was described at the IIIrd Brazilian Meeting
on Simulational Physics in 2003.  In Wang--Landau (WL) sampling, the
\emph{a priori} unknown density of states $g(E)$ of a system is
determined iteratively by performing a random walk in energy space
($E$) and sampling configurations with probability $1/g(E)$ (i.e. with
a ``flat histogram'')~\cite{wl_prl,wl_pre,shanho04ajp,3rdbmsp}. This
procedure has proven very powerful for studying wide ranging problems
with complex free energy landscapes because it circumvents the long
time scales typically encountered near phase transitions or at low
temperatures. The method also facilitates the calculation of
thermodynamic quantities, including the free energy, at any
temperature from a \emph{single} simulation.  Wang--Landau sampling is
also a generic Monte Carlo procedure with only a few adjustable
parameters, and it has been applied successfully to quite diverse
problems including spin glasses, polymers, protein folding, lattice
gauge theory, etc.,
see~\cite{rathore02jcp,alder04jsm,taylor09jcp,langfeld12prl} for
examples. Over the years, the method itself was subject to multiple
studies and various improvements to it have been
proposed~\cite{yamaguchi02pre,zhou05pre,wu05pre,zhou06prl,lee06cpc,belardinelli07pre,wuest09prl},
but the underlying simplicity remains intact.

A few simple attempts at parallelization of the WL algorithm have been
undertaken, but these are useful only for a relatively small number of
processors (cores).  One early approach~\cite{wl_pre,shanho04ajp} was
to subdivide the total energy range into smaller sub-windows, each
sampled by an independent WL random walker. The total simulation time
is limited by the convergence of the slowest walker, but it can be
tuned somewhat by an unequal partition of energy space. Nevertheless,
an optimal load balancing is impossible due to the \emph{a priori}
unknown irregularities in the complex free energy landscape, and the
individual energy intervals cannot be reduced arbitrarily due to
systematic errors introduced because some regions of
configurational space then become inaccessible.

An alternative scheme is to have multiple random walkers work
simultaneously on the \emph{same} density of states (and histogram).
Although this approach seemingly avoids some problems, a~recent,
massively parallel implementation~\cite{yin12cpc} revealed that
correlations among the walkers could lead to a systematically
underestimation of $g(E)$ in ``difficult to access'' energy regions.
A~proposed remedy to this problem was to add an ``ad hoc'' bias to the
modification factor; but such inter-dependencies are highly
problematic; besides the fact that an appropriate bias is, again,
\emph{a priori} unknown. It is also important to note that the
effective round-trip times of the individual walkers are not improved
by such an approach.

\section{The Replica Exchange Wang--Landau algorithm}

\begin{figure}[h!]
\centering
\begin{minipage}[b]{0.55\textwidth}
  \includegraphics[width=\textwidth,clip]{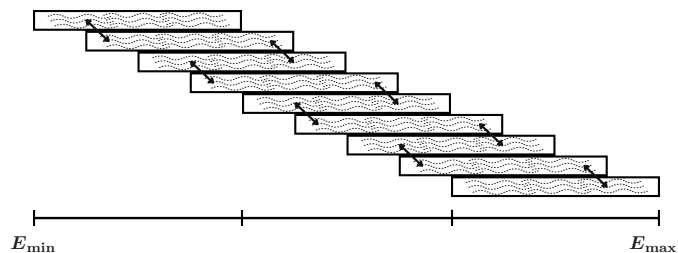}
\end{minipage}\hfill
\begin{minipage}[b]{0.35\textwidth}
  \caption{\label{fig:1}%
    Example of a splitting of the global energy range into nine
    equal-size intervals with a constant overlap of $75\%$. There are
    nine walkers in each energy interval, arrows indicate the generic
    replica-exchange path.}
\end{minipage}
\end{figure}

Our new approach~\cite{vogel13prl} is a \emph{generic} parallel method
which combines the advantageous dynamics of the original Wang--Landau
sampling scheme with the idea of replica-exchange Monte
Carlo~\cite{partemp1,partemp3}.  We begin by splitting up the total
energy range into $h$ smaller sub-windows with large overlap between
adjacent windows. The extent of the overlap $o$ should be chosen to
strike a balance between fast convergence of $g(E)$ and a reasonable
exchange acceptance rate; an overlap of, e.g., $o\approx75\%$ is a
good choice~\cite{vogel13pre}. See Fig.~\ref{fig:1} for a
visualization of a~corresponding generic setup. Each energy sub-window
is sampled by multiple ($m$), \emph{independent} WL walkers.  The key
to this approach is that configurational or replica exchanges are
allowed among WL instances of overlapping energy windows during the
course of the simulation.  As a~consequence, each replica can travel
through the entire energy space.  The replica exchange move does not
bias the overall WL procedure and, thus, guarantees the flexibility to
apply it to any valid WL update/convergence rule (e.g., the $1/t$
algorithm~\cite{belardinelli07pre}).  Furthermore, our hierarchical
parallelization approach imposes \emph{no} principal limitation to the
number of WL walkers or energy ranges so that it is possible to design
setups which scale up to many thousands of~CPUs.

\noindent
The standard WL algorithm~\cite{wl_prl,wl_pre} estimates the density
of states, $g(E)$, in an energy range $\left[ E_{\textrm{min}},
  E_{\textrm{max}}\right]$ using a single random walker. During the
simulation, trial moves are accepted with a probability $P =
\textrm{min} \left[1, g(E_\textrm{old})/g(E_\textrm{new})\right]$,
where $E_\textrm{old}$ ($E_\textrm{new}$) is the energy of the
original (proposed) configuration. The estimation of $g(E)$ is
continuously adjusted and improved using a modification factor $f$ (as
$g(E)\to f\times g(E)$) which progressively approaches unity as the
simulation proceeds, while a histogram $H(E)$ keeps track of the
number of visits to each energy $E$ during a given iteration.  When
$H(E)$ is sufficiently ``flat'', the next iteration begins with $H(E)$
reset to zero and $f$ reduced by some predefined rule (e.g. $f
\rightarrow \sqrt f$). The simulation terminates when $f$ reaches a
small enough $f_\textrm{final}$ at which point the accuracy of $g(E)$
is proportional to $\sqrt f_\textrm{final}$ for flat enough
$H(E)$~\cite{zhou05pre}.

In our parallel WL scheme, each random walker performs standard WL
sampling in its energy sub-window.  After a certain number of Monte
Carlo steps, a ``replica exchange'' is proposed between two random
walkers, $i$ and $j$, where walker $i$ chooses swap partner $j$ from a
neighboring sub-window at random. Let $X$ and $Y$ be the
configurations that the random walkers $i$ and $j$ are carrying before
the exchange; $E(X)$ and $E(Y)$ be their energies, respectively.  From
the detailed balance condition the acceptance probability $P_{acc}$
for the exchange of configurations $X$ and $Y$ between walkers $i$ and
$j$ is:
\begin{equation}
  P_{acc}=\min\left[1,\frac{g_i(E(X))}{g_i(E(Y))}\frac{g_j(E(Y))}{g_j(E(X))}\right]\,,
\end{equation}
where $g_i(E(X))$ is the instantaneous estimator for the density of
states of walker $i$ at energy $E(X)$, cf. also~\cite{nogawa11pre}.

\begin{figure}[b!]
\centering
\makebox[0.05\textwidth]{}
\begin{minipage}[b]{0.5\textwidth}
\includegraphics[width=\textwidth]{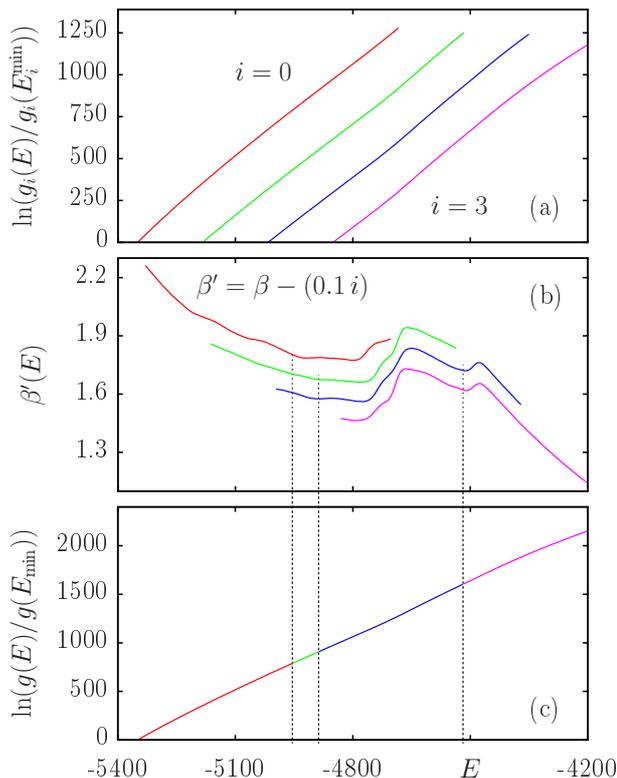}
\end{minipage}\hfill
\begin{minipage}[b]{0.35\textwidth}
  \caption{\label{fig:2}%
    Illustration of the process of connecting density of states pieces
    after the simulation. (a) Individual pieces from four overlapping
    energy sub-windows (labeled $i$). (b) The inverse microcanonical
    temperature $\beta(E)=$\break$\mathrm{d}\ln[g(E)]/\mathrm{d}E$.
    Grid lines indicate the positions where neighboring data coincide
    best; individual curves are shifted with respect to each other for
    clarity. (c) Connected global density of states. The color codes
    show which part originates from which energy sub-window.}
\end{minipage}
\end{figure}
An important new feature of our formalism is the provision of every
random walker with its own $g(E)$ and $H(E)$ which are updated
individually. Also, each walker must fulfill the WL flatness criterion
\emph{independently} at each iteration, ensuring that systematic
errors as found in~\cite{yin12cpc} cannot occur.  When \emph{all}
random walkers within an energy sub-window have attained flat
histograms, their estimators for $g(E)$ are averaged and redistributed
among themselves before simultaneously proceeding to the next
iteration.  This procedure reduces the statistical error during the
simulation with ${\sqrt{m}}$~\cite{vogel13pre}, i.e. as for
uncorrelated WL simulations. Furthermore, increasing $m$ can improve
the convergence of the WL sampling by reducing the risk of statistical
outliers in $g(E)$ which would slow down subsequent iterations.
Alternatively, it allows us, in principle, to use a weaker flatness
criterion, leading to an additional potential
speed-up~\cite{vogel13pre}.

The simulation terminates when all $h$ energy intervals have attained
$f_\textrm{final}$, and we are left with $h$ fragments of $g(E)$ with
overlapping energy intervals, see Fig.~\ref{fig:2}a for an example.
The pieces are then used to calculate a single $g(E)$ over the entire
energy range with the joining point for any two overlapping density of
states pieces chosen where the inverse microcanonical temperatures
$\beta=\textrm{d}\log[g(E)]/\textrm{d}E$ best coincide (see
Fig.~\ref{fig:2}\,b).  This guarantees that there are no
non-differentiable points in the microcanonical entropy $\log[g(E)]$,
which would lead to artificial peaks in observables like the heat
capacity. We set neighboring density-of-state pieces equal at those
joining points and cut away superfluous parts to yield a continuous
density of states, see Fig.~\ref{fig:2}\,c. To estimate statistical
errors, we repeat the simulation a few times and apply resampling
methods (bootstrapping)~\cite{newman_book}. Note that for each
resampled global density of states, the joining points between
neighboring pieces of data are at different positions.

\section{How well does the algorithm work?}

In order to assess the generality and performance of this novel
parallel WL scheme, we applied it to multiple intrinsically different
models.  We shall concentrate on two examples here: The well studied
10-state Potts model in 2 dimensions showing a strong first-order
transition, and a continuum model of amphiphilic molecules in solution
showing multiple structural transitions during the lipid bilayer
formation.

\subsection{The 10-state Potts model and weak scaling}

The Hamiltonian for the 10-state Potts model is given by
\begin{equation}
  \label{eq:potts}
  \mathcal{H}=-\sum_{\langle i,j \rangle} \delta(\sigma_i,\sigma_j)\,,
\end{equation}
where the $\sigma_i$ are spin variables which can take ten different
(integer) values, and ${\langle i,j \rangle}$ refers to all nearest
neighbor spin pairs. $N=L\times L$ is the total number of spins, $L$
the linear lattice size.

\begin{figure}[b]
\centering
  \includegraphics[width=0.9\textwidth,clip]{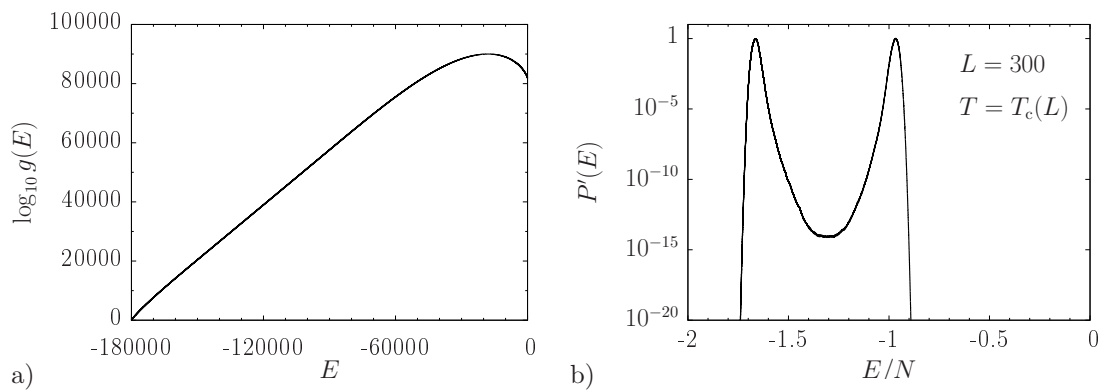}
  \caption{\label{fig:3}%
    Data for the $L=300$ 2D 10-state Potts model. (a) Logarithm of the
    density of states obtained by our parallel WL scheme with equal-size
    sub-windows. (b)
    Probability density (normalized to $P{'}=1$ at the peaks) at the
    critical temperature.  Transition states are suppressed by a
    factor of $\approx10^{-14}$.}
\end{figure}

Using single spin flip updates, serial
generalized-ensemble Monte Carlo methods are able to study system
sizes of about $N=200^2$ within reasonable time (order of
weeks)~\cite{wl_pre}. Figure~\ref{fig:3} shows the density of states
for the $N=300^2$ lattice, which we obtained in a few hours by
employing $\approx 2000$ parallel walkers (the energy range is split
into $717$ sub-windows with an overlap of 75\%, three walkers are
deployed in each window), and the respective canonical distribution at
the finite size transition temperature. The density of states covers
almost $100\,000$ orders of magnitude and transition states at the
phase transition are suppressed by a factor of $\approx10^{-14}$.
The finite-size transition temperature $T_\mathrm{c}(L)$ is measured
to be $0.70126$. We estimate the critical temperature in the
thermodynamic limit from results of system sizes up to
$300^2$~\cite{vogel13pre}. Our extrapolated value of
$T_\mathrm{c}^\infty=0.701234\pm0.000006$ is in excellent agreement
with the exactly known $T_\mathrm{c}=0.701232$~\cite{baxter73jpc}.

\begin{figure}
\centering
\makebox[0.05\textwidth]{}
\begin{minipage}[b]{0.5\textwidth}
  \includegraphics[width=\textwidth,clip]{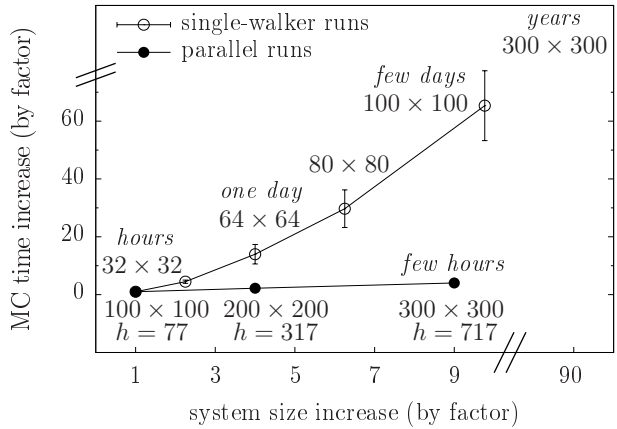}
\end{minipage}\hfill
\begin{minipage}[b]{0.35\textwidth}
\caption{\label{fig:4} Relative MC time to terminate serial WL runs
    for different system sizes for the 2D Potts model (open circles)
    and for parallel runs (filled circles). The number of energy
    windows $h$ in\-creases according to the increase in system size.}
\end{minipage}
\end{figure}

Besides this remarkable accuracy and absolute acceleration, our
algorithm shows very good weak scaling behavior for these lattice
models. This means that we can keep the simulation time almost
constant when making the system larger if we also increase the number
of computing cores by the same factor. In contrast, for serial runs
this time increases rapidly with increasing system size. See
Fig.~\ref{fig:4} for illustration, where we show the simulation times
for both serial and parallel runs.  Absolute run times are rough
estimates, $h$ is the number of energy sub-windows for parallel runs.

\subsection{Lipid bilayer formation in a continuous model}

\begin{figure}
\centering
  \includegraphics[width=0.67\textwidth,clip]{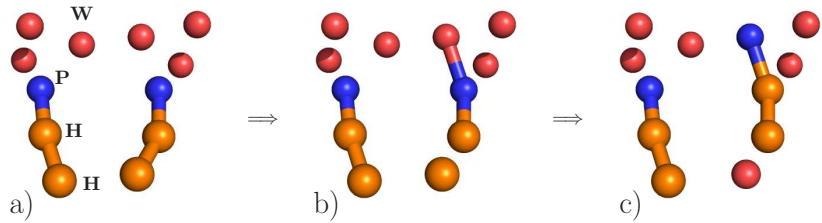}
  \caption{\label{fig:5}%
    Reptation trial move:  a) Amphiphilic molecules are composed of a polar (P)
    head bead and two hydrophobic (H) tail beads, and surrounded by
    solution particles (W). b) Intermediate step (not a valid configuration). c)
    Final reptation move configuration after
    particle reassignment.}
\end{figure}

To show how this parallel approach can open up new vistas we also
applied the method to a~very distinct and particularly challenging
molecular problem of high interest in biochemistry and pharmaceutical
science: a coarse-grained continuum model for the self-assembly of
amphiphilic molecules (lipids) in explicit solution. In this model,
amphiphilic molecules, each of which is composed of a polar (P) head
and two hydrophobic (H) tail monomers (P--H--H), are surrounded by
solvent particles (W), see Fig.~\ref{fig:5}\,a.  The interactions
between H and W molecules as well as those between H and P molecules,
are purely repulsive and given by the following potential:
\begin{equation}
  \label{eq:U_rep}
  U_\mathrm{rep}(r_{ij})=4\,\epsilon_\mathrm{rep}
  \left(\frac{\sigma_\mathrm{rep}}{r_{ij}}\right)^9\,,
\end{equation}
with $r_{ij}$ being the distance between two particles. All other
interactions between non-bonded particles are of Lennard-Jones type:
\begin{equation}
  \label{eq:U_LJ}
  U_\mathrm{LJ}(r_{ij})=
  4\epsilon\left[\left(\frac{\sigma}{r_{ij}}\right)^{12}
    -\left(\frac{\sigma}{r_{ij}}\right)^{6}\right]\,.
\end{equation}
As is usual when simulating generic coarse-grained models, we use
reduced units, i.e., set $k_\mathrm{B}=1$ and
$\epsilon_\mathrm{rep}=\epsilon_\mathrm{X-Y}=1$, where X-Y stands for
H-H, P-P, P-W, and W-W interactions. Furthermore, we set
$\sigma_\mathrm{rep}=1.05\sigma$ with $\sigma=1$,
cf.~\cite{getz,fujiwara} for similar models. Bonded monomers are
connected by a FENE-WCA~\cite{FENE,WCA} potential:
\begin{equation}
  \label{eq:U_WCA}
  U_\mathrm{FENE}(r_{ij})=
  -0.5KR^2\,\ln\left[1-\left(\frac{r_{ij}}{R}\right)^2\right]
  +U_\mathrm{LJ}(r_{ij})+\epsilon^\ast\,,
\end{equation}
where we set $K=30$ and $R=1.3$; and $\epsilon^\ast=\epsilon$ for
$r_{ij}<2^{1/6}\sigma$ or else $0$.

Such models offer qualitatively different technical challenges
compared to the Potts model and simulations of sufficiently large
setups to study the bilayer formation are \emph{impossible} for all
practical purposes with traditional, single walker Monte Carlo
methods. Besides local displacement moves, we introduce a reptation
move which facilitates creation (and persistance) of highly ordered
bilayer structures. During this move, a solution particle (W),
randomly chosen amongst all which are within a distance $R$ of either
the head or the tail of a given amphiphilic molecule, is ``captured''
and the respective other end of the lipid is ``released''~(Fig.~\ref{fig:5}\,b). To restore a valid configuration, particle
types are then adapted correspondingly (Fig.~\ref{fig:5}\,c). Note
that due to generally different numbers of ``close'' solution particles
at both ends of the lipid, forward and backward moves have different
selection probabilities, and this bias must be corrected in the Monte
Carlo acceptance probability.  We used a system containing $N=1000$
particles in total, including $M=125$ lipid molecules. One Monte Carlo
sweep consists of $N$ updates, on average $3M/10$ of which are
reptation moves. The size of the simulation box is chosen such that
the number density $\rho=0.8$.

\begin{figure}
\centering
  \includegraphics[width=0.9\textwidth,clip]{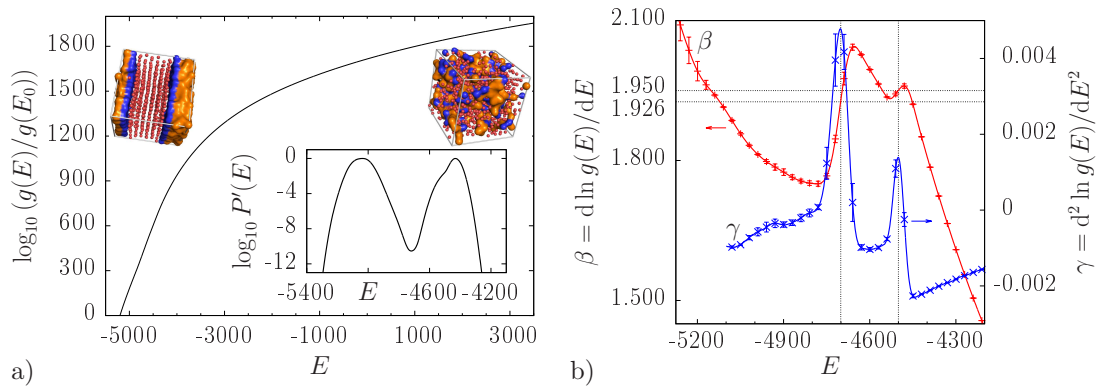}
  \caption{\label{fig:6}%
    Data for the $N=1000$-particle amphiphilic lipid model. (a)
    Logarithm of the density of states obtained with equal sub-windows
    as shown in Fig.~\ref{fig:1} and normalized probability density
    at the critical temperature.  Signs of a
    multi-stage transition can be seen in the right peak.  Inset
    pictures show example configurations with $E\approx-5350$ and
    $3000$, respectively. (b) First and second derivative of the
    microcanonical entropy (``microcanonical analysis'').   See text for details.}
\end{figure}
The density of states $g(E)$ over an energy range covering the
complete lipid bilayer formation from random solutions to frozen
bilayers spans $\approx2000$ orders of magnitude (Fig.~\ref{fig:6}\,a)
and cannot be sampled by single, serial Wang--Landau walkers in any
reasonable time. During the bilayer formation multiple transitions
occur, some almost simultaneously, as can be seen in the inset of
Fig.~\ref{fig:6}\,a where a dip in the right peak of the canonical
distribution is found. To understand such interlaced transitions we
examine the microcanonical inverse temperature~$\beta(E)$ and its
derivative~$\gamma(E)$ (see Fig.~\ref{fig:6}\,b.) There is a clear,
double back-bending in the inverse microcanonical
temperature~\cite{gross_book,junghans06prl}, indicating first-order
like transitions~\cite{schnabel11pre}, and two well separated peaks in
$\gamma(E)$. What we see here is a particular feature of the
microcanonical point of view: the pre-transition occurs at higher
energies, but lower (microcanonical) temperature.  It is important to
note that our algorithm does not miss such features.

\begin{figure}[b!]
  \makebox[0.05\textwidth]{}
  \includegraphics[width=0.7\textwidth,clip]{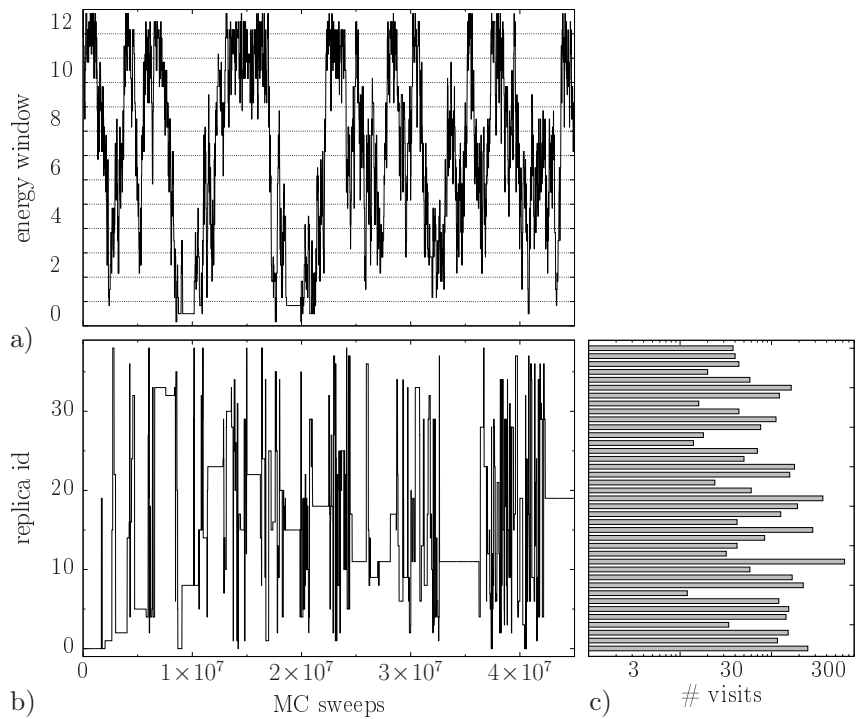}
  \caption{\label{fig:7}%
    a) Path of a single replica (contributing to data shown in
    Fig.~\ref{fig:6}) through the 13 energy sub-windows during the
    first $\approx 4\times 10^7$ MC sweeps. Replica exchanges are
    attempted every $10^4$ sweeps. b) and c) Data from the
    viewpoint of one of the WL walkers in the lowest-energy
    sub-window. We plot the ID of the replica which contributed to the
    histograms of that walker over time (b) and the corresponding
    histogram (c). All replicas contributed to the density of states
    estimator of that particular walker at some point (none of the
    histogram bins has zero entries).}
\end{figure}
Before we consider the physics of the lipid bilayer formation in
detail, we need to comment on another technical aspect of our method.
Due to the restricted local energy ranges, WL walkers are also
restricted to the corresponding parts of the phase space with the
potential of locking out parts of the overall phase space during the
simulation. The replica-exchange mechanism serves as a countermeasure
to this source of systematic error; hence, we must assure ourselves
that each replica can actually walk through the entire energy space.
As an example, in Fig.~\ref{fig:7}\,a we show the path of a replica
through energy sub-windows. In this setup, we split the energy range
into 13 overlapping sub-windows and deploy three WL walkers in each
window. At the beginning of the simulation one replica, with an energy
inside the walker's energy sub-window, is assigned to each of the 39
walkers. Hence, from another point of view, one can demand that every
WL walker, restricted to its small energy range, should be visited by
all replicas during the simulation. This is shown in
Fig.~\ref{fig:7}\,(b and c) for a walker in the lowest-energy window.
The histogram of visits (Fig.~\ref{fig:7}\,c) verifies that the walker
is actually visited by all replica (no histogram bin is empty).  Even
though this analysis does not guarantee that we covered all of phase
space during the simulation (in fact, one can never determine this),
the opposite behavior could indicate systematic errors.

Returning to the physical results, we first separate the energy scale
for the interaction between solution particles from all
others~\cite{gai13jcp}, to get a more unaltered look at the actual
transitions inside the lipid bilayer: we set $\epsilon_{W-W}=0.7$.
After estimating the density of states $g(E)$, we continued generating
data with fixed weights $1/g(E)$ and measure distributions $H(E,O)$
for geometrical observables $O$ to characterize the main structural
phases. We measure the following observables:

\paragraph{The bond order parameter $S_\mathrm{mol}^\mathrm{head}$~\cite{seelig74biochem}:}

\begin{equation}
  \label{eq:Smol}
  S_\mathrm{mol}^\mathrm{head}=1/2\,\left(3\langle\cos^2\theta\rangle-1\right)
\end{equation}
is a measure for the average orientation of bonds with respect to a
given direction in the simulation box. $\theta$ is the angle between a
bond connecting the head and center bead in a lipid and the x, y, or
z-direction, such that the order parameter is maximal~\footnote{Here
  and in the following we will refrain from doing principal component
  analyses of the amphiphilic cluster, as they are computationally
  very expensive, and use the principal axes of the simulation box
  instead. However, it turns out that low-energy structures will
  mostly align along those directions.}, and $\langle\cdot\rangle$
denotes the average over the respective bonds of all lipids. For
perfectly aligned bonds $S_\mathrm{mol}$ takes the maximum value of
$1$, its minimal value is $-0.5$, where all bonds are perpendicular to
a certain axis, e.g., for lipids forming a perfect cylinder and taking
the cylinder axis as reference. For spherical or random configurations
the parameter is close to $0$, independently of the reference axis.

\paragraph{Diagonal elements of the gyration tensor of the largest
  amphiphilic cluster and prolateness~\cite{aronovitz86jpa,blavatska10jcp}:}

To measure these quantities it is necessary to define and detect the
largest cluster of amphiphilic molecules in the system at every
measurement. This is efficiently done by a continuous implementation
of the Hoshen-Kopelman algorithm~\cite{hoshen76prb}. In analogy
to~\cite{fujiwara} two amphiphilic molecules are defined to be
neighbors whenever a tail monomer of one molecule is within a distance
$d<1.4\sigma$ from a tail monomer of the other molecule.  Furthermore,
as the extension of clusters is usually larger than the simulation box
size and in order to calculate meaningful quantities, all measurements
are done in unfolded coordinates, i.e., all pieces of a cluster
connected only via the periodic boundaries will be moved such that the
whole cluster is connected in free space. The diagonal elements of the
gyration tensor of the largest lipid cluster are calculated as
follows:
\begin{equation}
  \label{eq:Qii}
  Q_{ii}=\frac{1}{N}\sum_{j=1}^N\left(r^{(i)}_j-r_\mathrm{com}^{(i)}\right)\,,
\end{equation}
where $j$ labels an individual particle, $r$ is the position vector,
$i=1,2,3$ indicates one of the principal directions $x$, $y$, or $z$
and $r_\mathrm{com}^{(i)}$ is the position of the center of mass of
the unfolded cluster in that direction. We use $N$ here for the total
number of particles belonging to the cluster. We derive the
prolateness $P$ from the components of the gyration tensor in three
dimensions:
\begin{equation}
  \label{eq:prola}
  P=9\,\frac{\mathrm{Tr}\,\widetilde{Q}^3}{(\mathrm{Tr}\,Q)^3}\,,
\end{equation}
where $\mathrm{Tr}\,Q$ is the trace of the gyration tensor and
$\widetilde{Q}=Q-\lambda I$, with $\lambda=(\mathrm{Tr}\,Q)/3$ and $I$
being the unity matrix. The prolateness approaches its maximal value
of $2$ for thin cylinders or rod like structures ($Q_{11}\gg
Q_{22}\approx Q_{33}$) and its minimum value of $-1/4$ for disks
($Q_{11}\approx Q_{22}\gg Q_{33}\approx0$).

\begin{figure}
\centering
  \includegraphics[width=0.9\textwidth,clip]{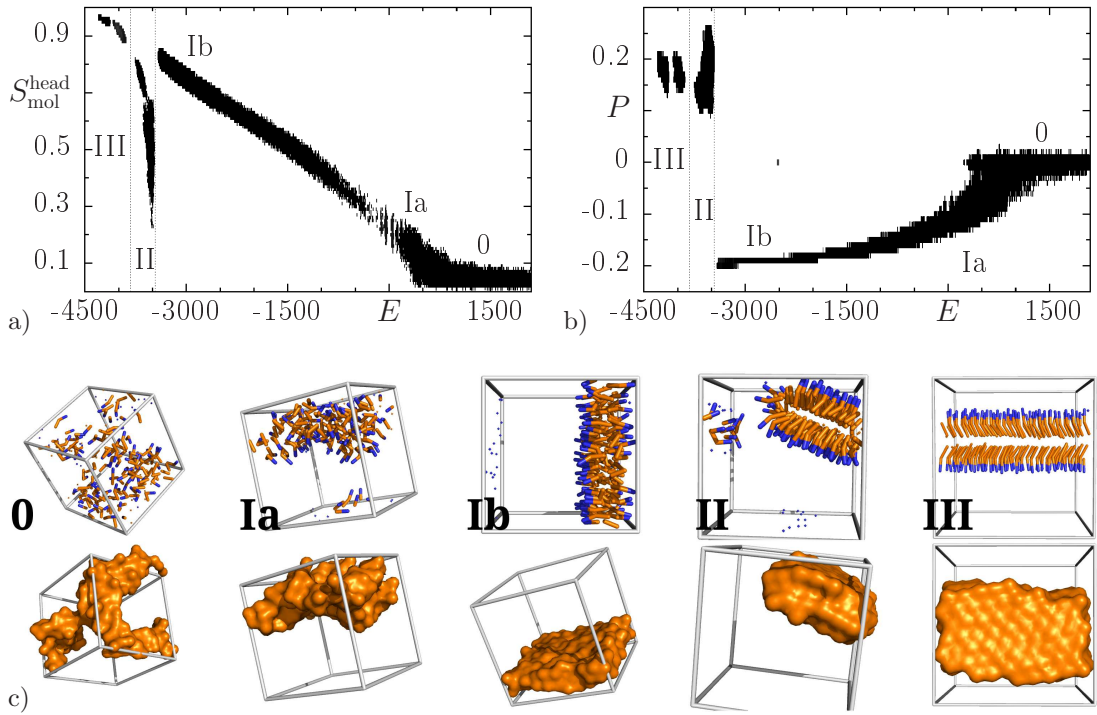}
  \caption{\label{fig:8}%
    Analysis of main structural phases during the bilayer formation.
    Top views on the two-dimensional energy--bond-order (a) and
    energy-prolateness histograms (b) measured during a WL production
    run. c) shows representative structures. Top row: single lipid
    representation, bottom row: cluster representation. Solution
    particles (W) are not shown for clarity.}
\end{figure}
Figure~\ref{fig:8} (a and b) shows top views of the histograms
$H(E,S_\mathrm{mol}^\mathrm{head})$ and $H(E,P)$. Data is combined
from all energy sub-windows, the grid-line positions correspond to
energies at which~$\gamma(E)$ (cf. Fig.~\ref{fig:6}\b) exhibits peaks.
For $E\gtrsim 0$ we see disordered, random solutions
($S_\mathrm{mol}^\mathrm{head}\approx P\approx 0$; phase\footnote{By
  ``phase'' we refer to phases within this particular finite system.}
``0''), at $E\approx 0$ cluster with a preferred orientation start to
form (``Ia''). Those clusters become more and more disc-like
($P\to-0.25$) and lipid molecules align
$S_\mathrm{mol}^\mathrm{head}\to1$ (``Ib''), but are still
``liquid-like''. Phase ``II'' can be seen as an intermediate phase
where two lipid layers separate and finally form a bilayer with
``crystallized'' lipids (phases ``III''). Figure~\ref{fig:8}\,c shows
representative structures from all phases with corresponding property
sets $(E;S_\mathrm{mol}^\mathrm{head};P)$. For clarity, solution
particles are not shown but were always included in simulations. For
a~comprehensive, more biochemical discussion of bilayer formation and
structural changes in the bilayer, see~\cite{gai13jcp}.

Such a complete thermodynamic analysis of a system of
this size, including explicit solution particles, was previously impossible
using Monte Carlo techniques. The proposed parallel
Wang--Landau method, however, enables us to enter a domain which had
been dominated by molecular dynamics simulations, bringing along
all the advantages of generalized ensemble MC techniques.

\section{Summary}

We have described a generic, parallel framework for Wang--Landau
simulations based on the concepts of energy range splitting, replica
exchange Monte Carlo, and multiple random walkers.  The method is
simple, generally applicable, and leads to significant advantages over
traditional, single-walker WL sampling.  Most importantly, it is
readily adaptable to parallel systems with an arbitrary number of
processors (cores) including systems with many thousands of
cores. As shown in two examples, our parallel WL framework allows us
to successfully simulate previously inaccessible systems.  The reasons
are two-fold: first, each walker is now responsible for sampling a
smaller configurational phase space, leading to 
faster convergence.  Second, the replica exchange process revitalizes
walkers in trapped states and avoids an erroneous bias in $g(E)$ due
to potential ergodicity breaking. We demonstrated
weak scaling behavior and the coverage of the entire energy space.
Analyses of the strong scaling abilities and the influence of
the number of walkers on the statistical error are given
elsewhere~\cite{vogel13prl,vogel13pre}.

\ack{This work was supported by the National Science Foundation under
  Grants DMR-0810223 and OCI-0904685. Y.W. Li was partly sponsored by
  the Office of Advanced Scientific Computing Research; U.S.
  Department of Energy. Supercomputer time was provided by TACC under
  XSEDE grants PHY130009 and PHY130014. Assigned: LA-UR-13-26563.}

\section*{References}

\begin{thebibliography}{99}

\bibitem{wl_prl} Wang F and Landau D P 2001 \textit{Phys. Rev. Lett.}
  \textbf{86} 2050

\bibitem{wl_pre} Wang F and Landau D P 2001 \textit{Phys. Rev.} E
  \textbf{64} 056101

\bibitem{shanho04ajp} Landau D P, Tsai S-H, and Exler M 2004
  \textit{Am. J.  Phys.} \textbf{72} 1294

\bibitem{3rdbmsp} Landau D P and Wang F 2003 \textit{Braz. J. Phys.}
  \textbf{34} 354

\bibitem{rathore02jcp} Rathore N and de Pablo J J 2002 \textit{J.
    Chem. Phys.}  \textbf{116} 7225

\bibitem{alder04jsm} Alder S, Trebst S, Hartmann A K, and Troyer M
  2004 \textit{J. Stat. Mech.} \textbf{2004} P07008

\bibitem{taylor09jcp} Taylor M P, Paul W, and Binder K 2009 \textit{J.
    Chem.  Phys.} \textbf{131} 114907

\bibitem{langfeld12prl} Langfeld K, Lucini B, and Rago A 2012
  \textit{Phys.  Rev. Lett.} \textbf{109} 111601

\bibitem{yamaguchi02pre} Yamaguchi C and Kawashima N 2002
  \textit{Phys. Rev.} E \textbf{65} 056710

\bibitem{zhou05pre} Zhou C and Bhatt R N 2005 \textit{Phys. Rev.} E
  \textbf{72} 025701(R)

\bibitem{wu05pre} Wu Y, K\"orner M, Colonna-Romano L, Trebst S, Gould
  H, Machta J, and Troyer M 2005 \textit{Phys. Rev.} E \textbf{72}
  046704
  
\bibitem{zhou06prl} Zhou C, Schulthess T C, Torbr\"{u}gge S, and
  Landau D P 2006 \textit{Phys. Rev. Lett.} \textbf{96} 120201

\bibitem{lee06cpc} Lee H K, Okabe Y, Landau D P 2006 \textit{Comp.
    Phys. Comm.} \textbf{175} 36

\bibitem{belardinelli07pre} Belardinelli R and Pereyra V 2007
  \textit{Phys. Rev.}  E \textbf{75} 046701

\bibitem{wuest09prl} W\"{u}st T and Landau D P 2009 \textit{Phys.
    Rev. Lett.} \textbf{102} 178101

\bibitem{yin12cpc} Yin J and Landau D P 2012 \textit{Comp. Phys.
    Comm.}  \textbf{183} 1568

\bibitem{vogel13prl} Vogel T, Li Y W, W\"ust T, and Landau D P 2013
  \textit{Phys. Rev. Lett.} \textbf{110} 210603

\bibitem{partemp1} Geyer C J 1991 \textit{Comp. Sci.  Stat.: Proc. of
    the 23rd Symp. on the Interface} ed Keramidas E M (Fairfax
  Station, VA: Interface Foundation) p 156

\bibitem{partemp3} Hukushima K and Nemoto K 1996 \textit{J. Phys. Soc.
    Jpn.}  \textbf{65} 1604

\bibitem{vogel13pre} Vogel T, Li Y W, W\"{u}st T, and Landau D P (in
  preparation)

\bibitem{nogawa11pre} Nogawa T, Ito N, and Watanabe H 2011
  \textit{Phys. Rev.} E \textbf{84} 061107 (note the misprint in the
  corresponding equation though)

\bibitem{newman_book} Newman M E J and Barkema G T 1999 \textit{Monte
    Carlo methods in statistical physics} (Oxford, New York: Oxford
  University Press)

\bibitem{baxter73jpc} Baxter R J 1973 \textit{J. Phys. C: Solid State
    Phys.} \textbf{6} L445

\bibitem{getz} Goetz R and Lipowsky R 1998 \textit{J. Chem. Phys.}
  \textbf{108} 7397

\bibitem{fujiwara} Fujiwara S, Itoh T, Hashimoto M, and Horiuchi R
  2009 \textit{J. Chem. Phys.} \textbf{130} 144901

\bibitem{FENE} Warner H R 1972 \textit{Ind. Eng. Chem. Fundam.}
  \textbf{11} 379

\bibitem{WCA} Chandler D, Andersen H C 1971 \textit{J. Chem. Phys.}
  \textbf{54} 26

\bibitem{gross_book} Gross D H E 2001 \textit{Microcanonical
    Thermodynamics} (Singapore: World Scientific)

\bibitem{junghans06prl} Junghans C, Bachmann M, and Janke W 2006
  \textit{Phys.  Rev. Lett.} \textbf{97} 218103

\bibitem{schnabel11pre} Schnabel S, Seaton D T, Landau D P, and
  Bachmann M 2011 \textit{Phys. Rev.} E \textbf{84} 011127

\bibitem{gai13jcp} Gai L, Vogel T, Maerzke K A, Iacovella C R,
  Landau D P, Cummings P T, and McCabe C 2013 \textit{J. Chem. Phys.}
  \textbf{139} 054505 

\bibitem{seelig74biochem} Seelig A and Seelig J 1974
  \textit{Biochemistry} \textbf{13} 4839

\bibitem{aronovitz86jpa} Aronovitz J A and Stephen M J 1987 \textit{J.
    Phys.  A: Math. Gen.} \textbf{20} 2539

\bibitem{blavatska10jcp} Blavatska V and Janke W 2010 \textit{J. Chem.
    Phys.}  \textbf{133} 184903

\bibitem{hoshen76prb} Hoshen J and Kopelman R 1976 \textit{Phys. Rev.}
  B \textbf{14} 3438

\end{thebibliography}

\end{document}